\documentclass[12pt,twoside,aps,pra,nofootinbib,amsmath,amssymb,tightenlines,floats,floatfix]{revtex4}
\usepackage{epsfig}
\usepackage{psfig}
\usepackage{mathptmx}
\usepackage{bm}
\renewcommand{\sl}{\rm}
\newcommand{\ral}{\rangle}
\newcommand{\lal}{\langle}
\newcommand{\Img}{\mathop{\rm Im}}
\newcommand{\Real}{\mathop{\rm Re}}
\newcommand{\D}{\displaystyle}

\newcommand{\cD}{{\mathcal D}}
\newcommand{\cG}{{\mathcal G}}
\newcommand{\cH}{{\mathcal H}}

\newcommand{\re}{{\rm e}}
\newcommand{\ri}{{\rm i}}
\newcommand{\sP}{{\sf P}}

\newcommand{\tM}{{\widetilde{M}}}

\newcommand{\reduction}[2]{\left.\phantom{\bigl|} #1 \right|_{#2}}
\newcommand{\C}{{\mathbb C}}
\newcommand{\R}{{\mathbb R}}

\begin{document}

\title{Perturbation of finite-lattice spectral levels \\
by nearby nuclear resonances\footnote{This work was supported by
the International Science \& Technology Center (ISTC) under the
project \# 1471.}\footnote{Published in {\it Particles and Nuclei,
Letters} {\bf 1 [118]} (2004), 15--24.}}
\author{A. K. Motovilov and V. B. Belyaev}
\address{Bogoliubov Laboratory of Theoretical Physics, JINR \\
141980 Dubna, Moscow Region, Russia}

\begin{abstract}
{\small We consider finite linear or cyclic crystalline structures with
molecular cells having narrow pre-threshold nuclear resonance. We prove that
if the real part of such a nuclear resonance lies within the energy band (the
convex hull of the energy levels) of the crystalline structure arising of a
separated molecular level, then there exist molecular crystalline states that
decay exponentially in time and the decay rate $\Gamma_R^{(m)}$ of these
states in the main order is described by the formula $\Gamma_R^{(m)}\cong
4\frac{\mathop{\rm Re} a}{\Gamma_R^{(n)}}$ where $a$ is the value of the
residue of the molecular channel transfer function at the nuclear resonance
point and $\Gamma_R^{(n)}$ is the nuclear resonance width.}

\bigskip

\end{abstract}

\maketitle

\thispagestyle{empty}

%%%%%%%%%%%%%%%%%%%%%%%%%%%%%%%%%%%%%%%%%%%%%%%%%%%%%%%%%%%%%%%%%%
\section{Introduction}
\label{Intro}
%%%%%%%%%%%%%%%%%%%%%%%%%%%%%%%%%%%%%%%%%%%%%%%%%%%%%%%%%%%%%%%%%%

Molecules are usually treated as purely Coulombic systems, while
the strong interaction between their nuclear constituents is
assumed to play a negligible role. However any Coulombic
molecular level lying above the lower threshold of the nuclear
subsystem, is embedded in the continuous spectrum of the nuclear
sub-Hamiltonian. The coupling between the molecular and nuclear
channels, hence, turns this level into a resonance (see, e.~g.,
Refs.~\cite{Albeverio72,Howland,Hunziker,Rauch,ReedSimonIV} and
references cited therein). Of course, due to the wide Coulombic
barrier between the nuclei and the short-range character of the
nuclear interaction, this coupling, and thus the width of the
resonance, which determines the fusion probability of the nuclear
constituents of the molecule, is in general extremely small.

However, as pointed out in \cite{BelMotTMF,BelMotEprint}, the
situation may be rather different if the nuclear subsystem of a
molecule has a sufficiently narrow near-threshold resonance.  Examples of
such nuclear systems  may be read off from the data presented in
\cite{ENSDF}.  Among them are even customary systems like
$p\,p\,{}^{16}$O and
$p\,{}^{17}$O~\cite{83Ajzenberg-Selove18-20,95Levels18-19},
i.\,e., the nuclear constituents of the water molecule $H_2{O}$
or the hydroxyl ion $OH^-$ with $O$ being the isotope
${}^{17}$O.  For $Li D$ and $H_2{O}$ the influence of
near-threshold nuclear resonances on the molecular properties
has been studied in \cite{BMS1,BMS2,BMS3} by estimating the
overlap integrals between the corresponding molecular and
nuclear wave functions.  The best known example of such
phenomena is the muon catalyzed fusion of deuteron and triton in
the $dt\mu$ molecule, where the near-threshold nuclear resonance
${}^5$He$(3/2^+)$ plays a decisive role~\cite{BreunlichOthers}.

Being motivated by the above special cases, we deal in the present
work like in \cite{BelMotTMF,BMS-JMP} with a rather general model
Hamiltonian related to the ones considered by Friedrichs
in~\cite{Friedrichs}.  This Hamiltonian consists of a nuclear
part, a molecular part with eigenvalues embedded in the
continuous spectrum of the nuclear part, and a weak coupling term
which turns these unperturbed eigenvalues into  molecular
resonances. Since the model is explicitly solvable, the mechanism
of formation of the resonances becomes clearly visible.

The following property pointed out
in~\cite{BelMotTMF,BelMotEprint} appears, in particular, as a
general feature:  if the nuclear channel itself has a narrow
resonance with a position close to the molecular energy,
then {\rm the width} (the imaginary part) {\rm of the resulting
molecular resonance is found to be inversely proportional to
the nuclear width}.  In other words, a large increase of the
decay rate of the molecular state, i.\,e. of the fusion
probability, is observed in this case.  Such a coincidence of
nuclear and molecular energies is, of course, a very rare
phenomenon in nature.

Influence of the narrow pre-threshold resonances on the
properties of infinite crystalline molecular structures was
studied in \cite{BMS-JMP}. In the present work we
concentrate on more realistic finite crystals. A goal of this
work is to show that the decay rate of a molecular state with
the energy close to a near-threshold resonance may be
considerably enhanced when arranging molecular clusters within
a finite crystalline structure. The reason is that in such a
configuration the original discrete molecular energy turns
into a set of energy levels. That is, even if the position of
the nuclear resonance differs from the original molecular
level, it can get within this set. This allows for a fine
tuning by exciting the crystalline structure to energies as
close as possible to the energy of the nuclear resonance.  We
show that the lattice states, which correspond to such an
initial choice of their quasimomentum distribution, decay
exponentially with a rate which is again inversely
proportional to the width of the nuclear resonance.

%%%%%%%%%%%%%%%%%%%%%%%%%%%%%%%%%%%%%%%%%%%%%%%%%%%%%%%%%%%%%%%%%%
\section{Two-channel molecular resonance model}
\label{twochannel}
%%%%%%%%%%%%%%%%%%%%%%%%%%%%%%%%%%%%%%%%%%%%%%%%%%%%%%%%%%%%%%%%%%

In this section we recall our main reasoning
\cite{BelMotTMF,BMS-JMP} regarding an influence of a near-threshold
nuclear resonance on the width of a molecular resonance in the
case of a single molecule.

%%%%%%%%%%%%%%%%%%%%%%%%%%%%%%%%%%%%%%%%%%%%%%%%%%%%%%%%%%%%%%%%%%
\subsection{Description of the model Hamiltonian}
\label{TwoChannelDescr}
%%%%%%%%%%%%%%%%%%%%%%%%%%%%%%%%%%%%%%%%%%%%%%%%%%%%%%%%%%%%%%%%%%

Let us consider a two-channel Hilbert space
$\cH=\cH_1\oplus\cH_2$ consisting of a nuclear
Hilbert space $\cH_1$ (channel 1) and a one-dimensional
molecular space $\cH_2=\C$ (channel 2).  The elements of
$\cH$ are represented as vectors
$
u=\left(\begin{array}{c}
       u_1 \\
       u_2
\end{array}\right)\,
$ where $u_1\in\cH_1$ and $u_2\in\cH_2$ ($u_2$ is simply a complex
number).  The inner product $\lal u,v\ral_{\cH}=\lal u_1,v_1\ral
+u_2\overline{v}_2$ in $\cH$ is naturally defined via the inner
products $\lal u_1,v_1\ral$ in $\cH_1$ and $u_2\overline{v}_2$ in
$\cH_2$.

As a Hamiltonian in $\cH$ we consider the $2\times2$
operator matrix
\begin{equation}
\label{H2}
A=\left(\begin{array}{cr}
h_1            & \quad b \\
\lal\,\cdot\,, b\ral   &        \lambda_2
\end{array}
\right),
\end{equation}
where $h_1$ is the (self-adjoint) ``nuclear Hamiltonian'' in
$\cH_1$, and $\lambda_2\in\R$  a trial molecular energy. A
vector $b\in\cH_1$ provides the coupling between the channels.
It should be mentioned that the Hamiltonian~(\ref{H2}) resembles
one of the well known Friedrichs models~\cite{Friedrichs}.

If there is no coupling between the channels, i.\,e. for
$b=0$, the spectrum of $A$ consists of the spectrum of
$h_1$ and the additional eigenvalue $\lambda_2$.  We
assume that the continuous spectrum $\sigma_c(h_1)$ of the
Hamiltonian $h_1$ is not empty and that the eigenvalue
$\lambda_2$  is embedded in $\sigma_c(h_1)$.  It is also assumed
that $\lambda_2$ is not a threshold point of $\sigma_c(h_1)$,
and that this spectrum is absolutely continuous in a
sufficiently wide neighborhood of $\lambda_2$.

A nontrivial coupling $(b\neq0)$ between the channels will, in
general, shift the eigenvalue $\lambda_2$ into an unphysical
sheet of the energy plane. The resulting perturbed energy
appears as a resonance, i.\,e.,  as a pole of the analytic (or,
more precisely, meromorphic) continuation of the resolvent
$r(z)=(A-z)^{-1}$  taken between suitable states (see,
e.\,g.,~\cite{ReedSimonIV}).  In the present work we assume
that such a continuation through the absolutely continuous
spectrum of $h_1$ in some neighborhood of $\lambda_2$ is
possible at least for the matrix element $\langle
r_1(z)b,b\rangle$ of the resolvent $r_1(z)=(h_1-z)^{-1}$. Then,
from the explicit representation for the resolvent $r(z)$
\cite{BelMotTMF,BMS-JMP}, one can easily see that the operator-valued
function $\reduction{P_2(A-z)^{-1}}{\cH_2}$ admits meromorphic
continuation to the same neighborhood, too.

The poles of $r(z)$ on the physical sheet are either
due to zeros of the {\it transfer function} (see \cite{MennMot})
$$
M_2(z)=\lambda_2-z-\beta(z)
$$
or due to poles of the
resolvent $r_1(z)$ (see \cite{BelMotTMF,BMS-JMP}). The latter correspond to the discrete
spectrum of the operator $h_1$ which may determine part of the
point spectrum of $A$. This is true, in particular, for the
multiple eigenvalues of $h_1$.  In any case it is obvious that
the perturbation of the eigenvalue $\lambda_2$ only corresponds
to solutions of the equation $M_2(z)=0$, i.\,e., of
\begin{equation}
\label{H2res}
z=\lambda_2-\beta(z).
\end{equation}
This equation has no roots $z$ with $\Img z\neq0$ on the
physical sheet.  Therefore, being eigenvalues of the self-adjoint
operator $A$, they have to be real. Thus, Eq.
(\ref{H2res}) may have solutions only on the real axis and in
the unphysical sheet(s) of the Riemann surface of the resolvent
$r_1(z)$.

We start with a brief discussion of the case where the
nuclear channel Hamiltonian $h_1$ generates no resonances
close to $\lambda_2$ in a domain $\cD$ of the unphysical sheet
which ajoins the physical sheet from below the cut.  This
assumption implies that for a wide set of unit vectors
$\widehat{b}=b/\|b\|$ the quadratic form $\beta(z)=\|b\|^2\lal
r_1(z)\widehat{b},\widehat{b}\ral$ can be analytically continued
in $\cD$.  Moreover, under certain smallness conditions for
$\|b\|$, Eq.~(\ref{H2res}) is uniquely solvable~\cite{MennMot}
in $\cD$ providing in the main order (see,
e.\,g.,~\cite{KMMM-YaF88,MotRemovalJMP})
\begin{equation}
\label{H2resonance}
z_2\mathop{=}\limits_{\|b\|\to 0}
\lambda_2-\lal r_1(\lambda_2+\ri0)b,b\ral
+o(\|b\|^2).
\end{equation}
The real and imaginary parts of the resonance
$z_2=E_R^{(2)}-\ri\D\frac{\Gamma_R^{(2)}}{2}$, thus, are given by
\begin{eqnarray}
\nonumber
E_R^{(2)} &=& \lambda_2-
\Real\lal r_1(\lambda_2+\ri0)b,b\ral+o(\|b\|^2), \\
\label{Gammaj}
\Gamma_R^{(2)} &=& 2\Img\lal r_1(\lambda_2+\ri0)b,b\ral
+o(\|b\|^2).
\end{eqnarray}

%%%%%%%%%%%%%%%%%%%%%%%%%%%%%%%%%%%%%%%%%%%%%%%%%%%%%%%%%%%%%%%%%%
\subsection{Perturbation of the molecular resonance
by a nearby nuclear resonance}\label{Perturbation}
%%%%%%%%%%%%%%%%%%%%%%%%%%%%%%%%%%%%%%%%%%%%%%%%%%%%%%%%%%%%%%%%%%

Our main interest concerns the opposite case of a
nuclear resonance
$z_1=E_R^{(1)}-\ri\D\frac{\Gamma_R^{(1)}}{2}$,
$\Gamma_R^{(1)}>0$, with a real part $E_R^{(1)}$ close to
$\lambda_2$. For the sake of simplicity we assume the
corresponding pole of $r_1(z)$ to be of first order.  Let the
element $b\in\cH_1$ be such that the function $\beta(z)$ admits
an analytic continuation into a domain $\cD$ which contains both
points $\lambda_2$ and $z_1$. This domain, moreover, is assumed
to belong to the unphysical sheet which adjoins the physical
sheet along the upper rim of the cut.  In $\cD$ the function
$\beta(z)$, thus, can be written as
\begin{equation}
\label{BetaRepres}
\beta(z)=\D\frac{a}{z_1-z}+\beta^{\rm reg}(z)
\end{equation}
with $\beta^{\rm reg}(z)$ being a holomorphic function.  For a
fixed ``structure function'' \hbox{$\widehat{b}=b/\|b\|$} we
have $|a|=C_a\|b\|^2$ with a constant $C_a$ determined by the
residue of $r_1(z)$ at $z=z_1$.  Note that this residue is
usually expressed in terms of resonance (Gamow)
functions (see for example~\cite{MotMathNach}).  In fact, we
assume that the resonance corresponds to an ``almost
eigenstate'' of $h_1$. That is, in principle  a limiting
procedure $\Gamma_R^{(1)}\to 0$ is possible so that
the resonance turns into a usual eigenvalue with an eigenvector
$\psi_1\in\cH_1.$  More precisely, we assume
\begin{equation}
\label{alimit}
C_a= C_a^{(0)}+o(1) \quad \mbox{as}\quad {\Gamma_R^{(1)}\to 0}
\end{equation}
with $C_a^{(0)}\equiv\lal\widehat{b},
\psi_1\ral\lal\psi_1,\widehat{b}\ral\neq 0$. This can be achieved,
e.\,g., if the Hamiltonian $h_1$ itself has a matrix
representation of the form~(\ref{H2}) and the resonance $z_1$ is
generated by a separated one-dimensional channel. In such a case
we would have $C_a^{(0)}=1$ (for details see
Ref.~\cite{BelMotEprint}, Sec.\,II).

Let
\begin{equation}
\label{ReaIma}
   \Real a>0 \quad\mbox{and}\quad\Img a\ll\Real a
\end{equation}
and, for $z\in\cD$,
$$
|\Img\beta^{\rm reg}(z)|\geq c_\cD\|b\|^2
\quad\mbox{and}\quad
|\beta^{\rm reg}(z)|\leq C_\cD \|b\|^2\,.
$$
with constants $c_\cD>0$ and $C_\cD>0$.  Furthermore,
the coupling between the channels in the
Hamiltonian~(\ref{H2}) is assumed to be so weak that
\begin{equation}
\label{H2Conditions}
\left|{\beta^{\rm reg}(z)}\right|\leq
C_\cD\|b\|^2\ll\Gamma_R^{(1)}\quad\mbox{while}\quad
|a|= C_a\|b\|^2\ll\left(\Gamma_R^{(1)}\right)^2.
\end{equation}
It can be expected that these conditions are fulfilled in
specific molecular systems even under the supposition that the
nuclear width $\Gamma_R^{(1)}$ itself is very small.

After inserting~(\ref{BetaRepres}) for $\beta(z)$,
Eq.~(\ref{H2res}) turns into the ``quadratic'' equation
$$
(\lambda_2-z)(z_1-z)-a+(z_1-z)\beta^{\rm reg}(z)=0
$$
which can be ``solved'', i.\,e., can be rewritten in form of two
equations
\begin{equation}
\label{H2roots}
z=\D\frac{\lambda_2+z_1-\beta^{\rm reg}(z)}{2}\pm
\sqrt{\left(\D\frac{\lambda_2-z_1-\beta^{\rm reg}(z)}{2}\right)^2+a}.
\end{equation}
By Banach's Fixed Point Theorem,  each of the
equations~(\ref{H2roots}) has only one solution in the domain
$\cD$.  In case of the sign ``$-$'' we denote the root
of~(\ref{H2roots}) by $z_{\rm nucl}$,  in case of the sign ``$+$''
by $z_{\rm mol}$.

According to \cite{BelMotTMF} and \cite{BMS-JMP}, the roots
$z_{\rm nucl}$ and $z_{\rm mol}$ of~(\ref{H2roots})
are essentially given by
\begin{eqnarray}
\label{zNucl}
 z_{\rm nucl}  & \cong & z_1 -
\D\frac{a}{\lambda_2-z_1-\beta^{\rm reg}(z_1)}\cong
z_1-\D\frac{a}{ \lambda_2-z_1 },\\
\label{zMol}
z_{\rm mol} & \cong & \lambda_2-\beta^{\rm reg}(\lambda_2+\ri0)
+\D\frac{a}{\lambda_2-z_1-\beta^{\rm reg}(\lambda_2+\ri0)}\cong
\lambda_2+\D\frac{a}{\lambda_2-z_1}.
\end{eqnarray}
{}From the second condition~(\ref{H2Conditions}) follows
$\left|\D\frac{a}{\lambda_2-z_1}\right|\ll \Gamma_R^{(1)}$.
Consequently, this term provides in $z_{\rm nucl}$ a very small
perturbation of the initial nuclear resonance $z_1$.  As
compared to $\Gamma_R^{(1)}$ it represents also in $z_{\rm mol}$
a very weak perturbation of the molecular energy
$\lambda_2$.  However, as compared to the
result~(\ref{H2resonance}), valid in case of a missing nearby
nuclear resonance, it can be rather large.  In particular,
if the molecular energy $\lambda_2$ coincides with the real
part $E_R^{(1)}$ of the nuclear resonance $z_1$, then
$z_{\rm mol}=E_R^{(m)}-i\D\frac{\Gamma_R^{(m)}}{2}$ with
\begin{equation}
\label{GmFin}
E_R^{(m)}\cong\lambda_2-2\D\frac{\Img a}{\Gamma_R^{(1)}}
\quad\mbox{and}\quad
\Gamma_R^{(m)}\cong 4\D\frac{\Real a}{\Gamma_R^{(1)}}.
\end{equation}
{\it The width of the molecular resonance $z_{\rm mol}$ in
the presence of a nearby nuclear resonance $z_1$, thus,
turns out to be inversely proportional to the nuclear width
$\Gamma_R^{(1)}$}.

The second inequality~(\ref{H2Conditions}), chosen as a
condition for $\|b\|$ reflects the fact that the ``usual''
molecular width $\Gamma_R^{(2)}$ is much smaller than the width
of a usual nuclear resonance $\Gamma_R^{(1)}$,
\begin{equation}
\label{G2G1}
C_a\Gamma_R^{(2)}\ll  c_\cD\left(\Gamma_R^{(1)}\right)^2.
\end{equation}
This can practically always be assumed for concrete molecules.

Under condition~(\ref{alimit}) the value of $C_a=|a|/\|b\|^2$
differs from zero,  $C_a\geq C>0$, as $\Gamma_R^{(1)}\to 0$.
Therefore, in the presence of a narrow
($\Gamma_R^{(1)}\ll\D{C_a}/{c_\cD}$) nuclear resonance close
to $\lambda_2$ the molecular width $\Gamma_R^{(m)}$ is much
larger than the molecular width $\Gamma_R^{(2)}$ observed in
absence of such a resonance.  In fact, this ratio is determined by
the large quotient $\D\frac{C_a/c_\cD}{\Gamma_R^{(1)}}$.

%%%%%%%%%%%%%%%%%%%%%%%%%%%%%%%%%%%%%%%%%%%%%%%%%%%%%%%%%%%%%%%%%%
\section{Molecular resonances in a finite crystalline lattice}
\label{Lattice1Dim}
%%%%%%%%%%%%%%%%%%%%%%%%%%%%%%%%%%%%%%%%%%%%%%%%%%%%%%%%%%%%%%%%%%

Let us assume that the ``molecules'' described by the
Hamiltonian~(\ref{H2}) are arranged in
form of an finite one-dimensional linear (chain) crystalline
structure.  To describe such a crystal we introduce the
lattice Hilbert space
\begin{equation}
\label{cG}
\cG=\mathop{\oplus}\limits_{i=1}^{n}\cH^{(i)}
\end{equation}
representing an orthogonal sum of the Hilbert spaces
associated with the individual cells
\begin{equation}
\label{cHi}
\cH^{(i)}=\cH^{(i)}_1\oplus\cH^{(i)}_2\,.
\end{equation}
Here the subspaces $\cH^{(i)}_1\equiv\cH_1$ and
$\cH^{(i)}_2\equiv\cH_2\equiv\C$ are exactly the same ones as in
Sec.~\ref{twochannel} and, thus, $\cH^{(i)}\equiv\cH$.
The elements of the total Hilbert space $\cG$ are represented by the
sequences
$u=(u^{(1)},u^{(2)},\ldots,u^{(n)})$
with components
$
u^{(i)}=\left(\begin{array}{c}
       u^{(i)}_1 \\
       u^{(i)}_2
\end{array}\right)
$
where $u^{(i)}_1\in\cH_1$ and $u^{(i)}_2\in\cH_2=\C$.
The inner product in $\cH$ is defined by
$
\lal u,v\ral_{\cH}=\sum\limits_{i=1}^{n}
\lal u^{(i)},v^{(i)}\ral_{\cH^{(i)}}. $ The subspaces
$\cG_1=\mathop{\oplus}\limits_{i=1}^{n}\cH^{(i)}_1$ and
$\cG_2=\mathop{\oplus}\limits_{i=1}^{n}\cH^{(i)}_2$, with
$\cG=\cG_1\oplus\cG_2$, represent respectively the pure
nuclear and pure molecular channels.

In the present section we will first deal with the Hamiltonian $H$
acting in $\cH$ according to
\begin{align}
\nonumber
(Hu)^{(1)}=&Au^{(1)}+Wu^{(2)}\\
\label{1dimLatHam}
(Hu)^{(i)}=&Wu^{(i-1)}+Au^{(i)}+Wu^{(i+1)},\quad i=2,\ldots,n-1\\
\nonumber
(Hu)^{(n)}=&Wu^{(n-1)}+Au^{(n)}
\end{align}
where only the interaction between neighboring cells
is taken into account and the interaction operator
$W$ is chosen in the simplest form
\begin{equation}
\label{W1dim}
W=\left(\begin{array}{ccc}
       0              & & 0 \\
       0              & & {\sl w}
\end{array}\right)\,
\end{equation}
with ${\sl w}$ being a positive number.  Such a
choice of the interaction corresponds to the natural assumption
that the cells interact between each other via the molecular
states, while the direct interaction between nuclear constituents
belonging to different cells is negligible.  We assume that the
closed interval \mbox{$[\lambda_2-2{\sl w},\lambda_2+2{\sl w}]$}
is totally embedded in the continuous spectrum $\sigma_c(h_1)$
of $h_1$ and, moreover, that no thresholds of
$\sigma_c(h_1)$ belong to this interval.
For the sake of simplicity we also
assume that the interval belongs to the domain $\cD$ introduced
in Sec.~\ref{twochannel} and that for any
\mbox{$\mu\in[\lambda_2-2{\sl w},\lambda_2+2{\sl w}]$}
\begin{equation}
\label{NeverZero}
\Img\lal r_0(\mu\pm\ri0)\widehat{b},\widehat{b}\ral\neq0.
\end{equation}

Obviously, the Hamiltonian~(\ref{1dimLatHam}) is a self-adjoint
operator on the domain
$\mathop{\rm Dom}(H)=\mathop{\oplus}\limits_{i=1}^{n}D^{(i)}$
with $D^{(i)}=\mathop{\rm Dom}(h_1)\oplus\C$. The resolvent $R(z)=(H-z)^{-1}$ of
$H$ possesses a natural block structure,
$R(z)=\{R(j,k;z)\}$, $j,k=1,2,...,n$.
The blocks $R(j,k;z)$ satisfy the equations
\begin{align}
\label{Rjk}
& WR(j-1,k;z)+(A-z)R(j,k;z)+WR(j+1,k;z)=\delta_{jk}I,\\
\nonumber
&\quad j,k=1,2,\ldots,n,
\end{align}
where $\delta_{jk}$ stands for the Kronecker delta and $I$ for
the identity operator in the Hilbert space $\cH$ of cells.
Hereafter we assume $\Img z\neq0$ so that the value of $z$
automatically belongs to the resolvent set of the operator $H$.
The blocks $R(j,k;z)$ themselves possess a $2\times2$ matrix
structure, $R(j,k;z)=\{R_{mn}(j,k;z)\}$, $m,n=1,2,$
corresponding to the decomposition
\mbox{$\cH=\cH_1\oplus\cH_2$}.

The set of the sequences $f_k$, $k=1,2,\ldots,n$, with the elements
\begin{align}
\label{fk}
f_k(j)=&\sqrt{\frac{2}{n+1}}\sin(p_k j),\\
\label{pk}
&\quad p_k=\frac{\pi k}{n+1},
\end{align}
forms an orthonormal basis for the ($n$-dimensional) Hilbert
space $l_2^{n}$ of $n$-element sequences of the form $\{x_1,x_2,\ldots,x_n\}$,
$x_j\in\mathbb{C}$, $j=1,2,\ldots,n$. The Fourier transform
\begin{equation}
\label{Fourier}
(Fu)(p_k)=\sqrt{\frac{2}{n+1}} \sum_{j=1}^{n}  u^{(j)}\sin(p_k j)
\end{equation}
in $\cG$ reduces Eq.~(\ref{Rjk}) to
\begin{equation}
\label{Rpp}
(A-z)R(p_k,p_{k'};z)+2\cos p_k\,WR(p_k,p_{k'};z)=
\delta_{kk'}\,I\,\quad k,k'=1,2,\ldots,n,
\end{equation}
and the numbers $R(p_k,p_{k'};z)$ represent the matrix
elements of the resolvent $R(z)$ in this representation.
{}From~(\ref{Rpp}) it immediately follows that
\begin{equation}
\label{RFactor}
R(p_k,p_{k'};z)=G(p_k;z)\delta_{kk'},
\end{equation}
where
\begin{equation}
\label{Gpz}
G(p;z)=\left(\begin{array}{ccc}
r_1(z)+\D\frac{r_1(z)b\lal\,\cdot\,,b\ral r_1(z)}{\tM_2(p;z)} &\, &
-\D\frac{r_1(z)b}{\tM_2(p;z)}\\
-\D\frac{\lal\,\cdot\,,b\ral r_1(z)}{\tM_2(p;z)}
&\,& \D\frac{1}{\tM_2(p;z)}
\end{array}\right).
\end{equation}
Here, the scalar function ${\tM_2(p;z)}$ reads
\begin{equation}
\label{TrFunctModif}
\tM_2(p;z)=\lambda_2-z+2{\sl w}\cos p-\beta(z)\,.
\end{equation}
The numbers $p_k$ given by (\ref{pk}) represent the quasimomenta
of the finite crystalline structure under consideration.

Consider now the time evolution of the system described by the
Hamiltonian $H$ starting from a pure molecular state
$\varphi=\varphi_1\oplus\varphi_2$, $\|\varphi_m\|\in\cG_m$,
$m=1,2$, with $\varphi_1=0$ and $\|\varphi\|=\|\varphi_2\|=1$.
The probability to find the system at a time
$t\geq0$ in the molecular channel is given by
\begin{equation}
\label{Pphi}
     P_{\rm mol}(\varphi,t)=\|\sP_2\re^{-\ri Ht}\varphi\|^2,
\end{equation}
where $\sP_2$ is the orthogonal projection in $\cG$
on the pure molecular subspace $\cG_2$.
Obviously, one can represent the time evolution
operator $\exp(-\ri Ht)$ in terms of the
resolvent $R(z)=(H-z)^{-1}$,
\begin{equation}
\label{intexp}
\exp(-\ri Ht)=-\D\frac{1}{2\pi\ri}\D\oint\limits_\gamma
dz \,{\rm e}^{-\ri zt}(H-z)^{-1},
\end{equation}
where the integration is performed along a counterclockwise
contour $\gamma$ in the physical sheet
encircling the spectrum of the Hamiltonian $H$.

According to Eqs.~(\ref{RFactor}) and~(\ref{TrFunctModif}) the
operator $\reduction{\sP_2 (H-z)^{-1}}{\cG_2}$ acts in quasi-momentum
representation as the multiplication operator,
\begin{equation}
\label{G22}
   \biggl(\sP_2 R(z)\varphi\biggr)(p_k)=
    \D\frac{1}{\tM_2(p_k;z)}\,\varphi_2(p_k).
\end{equation}
Here $\varphi_2(p_k)$ stands for the values of the
Fourier transform~(\ref{Fourier}) of the vector
$$
\varphi_2=(\varphi_2^{(1)},\varphi_2^{(2)},\ldots,\varphi_2^{(n)}).
$$
Hence
\begin{equation}
\label{P2phiFactor}
   \biggl(\sP_2\re^{-\ri Ht}\varphi\biggr)(p_k)=
       - \D\frac{1}{2\pi\ri}\,\, \varphi_2(p_k)\,J(p_k,t)
\end{equation}
with
\begin{equation}
\label{Jpt}
J(p_k,t)=\D\oint\limits_\gamma dz \,\D\frac{\exp(-\ri z t)}
       {\,\widetilde{\lambda}_2(p_k)-z-\beta(z)\,}\,.
\end{equation}
Repeating almost literally the analysis of Section III in
\cite{BMS-JMP} one finds that the asymptotics of the
term $J(p_k,t)$ as $t\to\infty$ reads as follows
\begin{eqnarray}
\nonumber
J(p_k,t) &=& \exp\{-{\rm i}z_{\rm mol}(p_k) t\}\\
\nonumber
&&\times\Biggl[1-\frac{a}{\bigl(\widetilde{\lambda}_2(p_k)-z_1-
\beta^{\rm reg}(\widetilde{\lambda}_2(p_k)+\ri0)\bigr)^2}
  +
O\biggl(\varepsilon^4(p_k,\widetilde{\lambda}_2(p_k)+\ri0)\biggr)\Biggr]  \\
\label{JptBehavior}
 &\quad\quad +& \exp\{-{\rm i}z_{\rm nucl}(p) t\} \\
\nonumber
&&\times\Biggl[\D\frac{a}{\bigl(\widetilde{\lambda}_2(p_k)-z_1-
\beta^{\rm reg}(z_1)\bigr)^2} +
O\bigl(\varepsilon^4(p_k,z_1)\bigr)\Biggr]
+\widetilde{\varepsilon}(p_k,t)\,,
\end{eqnarray}
where
\begin{equation}
\label{epsilonpz}
\varepsilon(p_k,z)=\D\frac{a}
{[\widetilde{\lambda}_2(p_k)-z_1-\beta^{\rm reg}(z)]^2}\,.
\end{equation}
The function $\widetilde{\varepsilon}(p,t)=O(\|b\|^2)$ is always
small, $|\widetilde{\varepsilon}(p,t)|\ll1$.
By (\ref{zNucl}) and (\ref{zMol}) for the positions
of the resonance poles we then obtain
\begin{eqnarray}
\label{zNuclp}
 z_{\rm nucl}(p_k)  & \cong &
z_1-\D\frac{a}{ \lambda_2+2{\sl w}\cos p_k-z_1 },\\
\label{zMolp}
z_{\rm mol}(p_k) & \cong &
\lambda_2+2{\sl w}\cos p_k+\D\frac{a}{\lambda_2+2{\sl w}\cos p_k-z_1}.
\end{eqnarray}

The asymptotics~(\ref{JptBehavior}) implies
\begin{equation}
\label{IntExpGamma}
P_{\rm mol}(\varphi,t)=\sum_{k=1}^{n}
|J(p_k,t)|^2\,|\varphi_2(p_k)|^2=
\sum_{k=1}^{n}
\exp\{-{\Gamma_R^{(m)}(p_k)}\,t\}\,|\varphi_2(p_k)|^2
+\widetilde{\varepsilon}(t)\,
\end{equation}
where
\begin{equation}
\label{GammaRm}
\Gamma_R^{(m)}(p_k)=
-2\Img z_{\rm mol}(p_k)\cong
-2\Img\D\frac{a}{\lambda_2+2{\sl w}\cos p_k-z_1}.
\end{equation}
The background term $\widetilde{\varepsilon}(t)$
in~(\ref{IntExpGamma}) is small for any $t\geq 0$,
$\widetilde{\varepsilon}(t)=O(\|b\|^2)$ and
$|\widetilde{\varepsilon}(t)|\ll1$.

Further, let us assume that the number $n$ of cells in the
lattice is large and the real part $E_R^{(1)}$ of the nuclear
resonance $z_1$ belongs to the interval $[\lambda_2-2{\sl
w},\lambda_2+2{\sl w}]$, that is
$|E_R^{(1)}-\lambda_2|\leq2{\sl w}$. Then, one can always
prepare an initial molecular state $\varphi$ which decays via
the nuclear channel with a rate close to $4\D\frac{\Real
a}{\Gamma_R^{(1)}}$ (cf. formula~(\ref{GmFin})). Indeed, under the
assumption~(\ref{ReaIma}), this maximum is given by
$$
\mathop{\rm max}
\limits_{0\leq p\leq\pi}
\Gamma_R^{(m)}(p)\cong 4\D\frac{\Real a}{\Gamma_R^{(1)}}.
$$
The most appropriate is the monochromatic molecular state
$\varphi$ with the only nonzero component $\varphi_2(p_{k_0})$
associated with the quasimomentum $p_{k_0}$ closest to
$$
  p_{\rm max}=\mathop{\rm arccos}\D\frac{E_R^{(1)}-\lambda_2}{2{\sl w}}\,.
$$
In particular, if the values of $\varphi_2(p_k)$ are nonzero only for
quasimomenta $p_k$ restricted by
$$
\left|\cos
p_k-\D\frac{E_R^{(1)}-\lambda_2}{2{\sl w}}\right|
\leq\delta\,\D\frac{\Gamma_R^{(1)}}{4{\sl w}}
$$
with some small
$\delta>0$, then the width $\Gamma_R^{(m)}$ given by the
relation~(\ref{GammaRm}) varies in an interval lying
approximately between $\D\frac{1}{1+\delta^2}\, \D\frac{4\Real
a}{\Gamma_R^{(1)}}$ and $\D\frac{4\Real a}{\Gamma_R^{(1)}}$.

In a similar way one also treats a one-dimensional cyclic crystalline
structure. In this case the Hilbert space $\cG$ is the same as in
(\ref{cG}) but the operator $H$ reads
\begin{align}
\nonumber
(Hu)^{(1)}=&Wu^{(n)}+Au^{(1)}+Wu^{(2)}\\
\label{Hcyclic}
(Hu)^{(i)}=&Wu^{(i-1)}+Au^{(i)}+Wu^{(i+1)},\quad i=2,\ldots,n-1\\
\nonumber
(Hu)^{(n)}=&Wu^{(n-1)}+Au^{(n)}+Wu^{(1)}.
\end{align}
If the intercellur interaction is still given by (\ref{W1dim}), the
only difference in the analysis will be the use of another
complete orthonormal set in the space $l_2^{n}$. Instead of
the sequences (\ref{fk}) one now employs the ortonormal sequences
\begin{align}
\label{fkk}
f_k(j)=&\frac{1}{\sqrt{n}}\exp(\ri p_k j),\quad j=1,2,\ldots,n
\end{align}
with quasimomenta $p_k$ given by
\begin{equation}
\label{pkk}
p_k=\frac{2\pi k}{n}, \quad k=1,2,\ldots,n.
\end{equation}
After the Fourier transform (\ref{Fourier}) with
$\frac{1}{\sqrt{n}}\exp(\ri p_k j)$ the matrix elements of the
resolvent $(H-z)^{-1}$ again acquire the form (\ref{RFactor}),
(\ref{Gpz}) with the transfer function $\tM_2(p;z)$ given by
(\ref{TrFunctModif}). Hence, one concludes with formulas like in
(\ref{P2phiFactor})--(\ref{GammaRm}) and then observes that if the
number of cells is large enough it is possible to prepare
pure molecular states that decay with the rate close
to $4\frac{\Real a}{\Gamma_R^{(1)}}$.

In the same way one can also consider the finite two- and
three-dimensional crystalline structures arranged of the
molecular cells described by the Hamiltonian (\ref{H2}).
If the cell has a sharp near-threshold nuclear resonance
with energy embedded into the convex hull of the arising
crystalline molecular levels, one again will find an
enhancement of the decay rate for particular molecular
states. As in the case of the one-dimensional lattices
these molecular states should decay with the rate close to
$4\frac{\Real a}{\Gamma_R^{(1)}}$.

%%%%%%%%%%%%%%%%%%%%%%%%%%%%%%%%%%%%%%%%%%%%%%%%%%%%%%%%%%%%%%%%%
%%%%%%%%%%%%%%%%%%%%%%%%%%%%%%%%%%%%%%%%%%%%%%%%%%%%%%%%%%%%%%%%%
%%%%%%%%%%%%%%%%%%%%%%%%%%%%%%%%%%%%%%%%%%%%%%%%%%%%%%%%%%%%%%%%%

\newpage

\thispagestyle{empty}

\centering
\epsfig{file=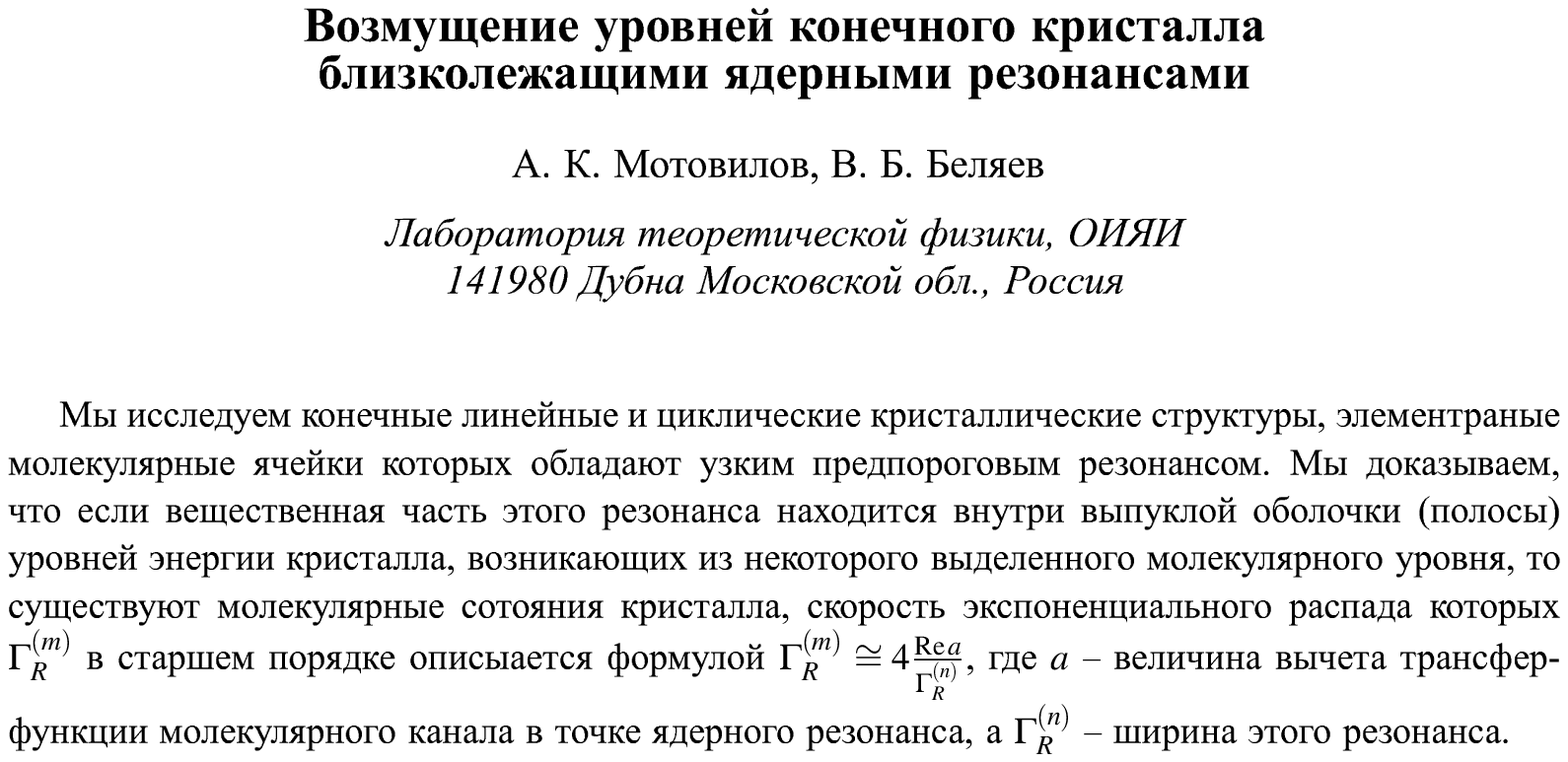,width=16.0truecm}

\end{document}